\documentclass[review,number,sort&compress]{elsarticle}

\usepackage{amsmath,color}
\usepackage{graphicx}

\usepackage{hyphenat}

\DeclareMathOperator{\sinc}{sinc}

\title{On the resolution of a MIEZE spectrometer}
\author{N. Martin}\address{Laboratoire L\'eon Brillouin, CEA, CNRS, Universit\'e Paris-Saclay, CEA Saclay 91191 Gif-sur-Yvette, France}\ead{nicolas.martin@cea.fr}
\date{\today}

\begin{document}

\begin{abstract}
We study the effect of a finite sample size, beam divergence and detector thickness on the resolution function of a MIEZE spectrometer. We provide a transparent analytical framework which can be used to determine the optimal trade-off between incoming flux and time-resolution for a given experimental configuration. The key result of our approach is that the usual limiting factor of MIEZE spectroscopy, namely neutron path length differences throughout the instrument, can be suppressed up to relatively large momentum transfers by using a proper small-angle (SANS) geometry. Under such configuration, the hitherto accepted limits of MIEZE spectroscopy in terms of time-resolution are pushed upwards by typically an order of magnitude, giving access to most of the topical fields in soft- and hard-condensed matter physics.
\end{abstract}

\maketitle
%%%%%%%%%%%%%%%%%%%%%%%%%%%%%%%%%%%%%%%%%%%%%%%%%%%%%%%%
%%%%%%%%%%%%%%%%%%%%%%%%%%%%%%%%%%%%%%%%%%%%%%%%%%%%%%%%
%%%%%%%%%%%%%%%%%%%%%%%%%%%%%%%%%%%%%%%%%%%%%%%%%%%%%%%%
%%%%%%%%%%%%%%%%%%%%%%%%%%%%%%%%%%%%%%%%%%%%%%%%%%%%%%%%
A majority of scientific advances are driven by technical development and the topics covered by neutron spectroscopy do not escape this paradigm. Constant efforts aiming at an improvement of momentum (space) or energy (time) resolution are crucial for addressing modern issues in soft- and hard-condensed matter physics. To date, the technique offering the finest energy resolution is Neutron Spin Echo (NSE) spectroscopy which allows studying slow processes ({\it i.e.} with characteristic times approaching the $\mu$s range), provided that the carefully manipulated beam polarization is not degraded by the sample or its environment \cite{Schleger1999}. Here we consider a derivative of NSE, the so-called MIEZE technique, where all spin manipulations are performed upstream of the sample position. At equivalent technical resolution, MIEZE is potentially more versatile than NSE since it works with any kind of samples ({\it e.g.} hydrogen-containing systems \cite{Besenboeck1998}, multi-domain ferromagnets \cite{Kindervater2017}, \emph{etc.}) and under extreme conditions ({\it e.g.} large magnetic fields \cite{Kindervater2015}). On the downside, being a time-of-flight technique, the efficiency of the method is limited towards high resolution by deviations from the optimal neutron flight path across the setup. Here, we show that these limitations can be drastically softened by using MIEZE in a small-angle (SANS) configuration. Our findings clearly pledge for the construction of a dedicated MIEZE-SANS instrument. Its performances can be quantified using the analytical framework developed in this paper. 

\section{Principles and limits of MIEZE spectroscopy}

\begin{figure}[!ht]
	\includegraphics[width=0.98\textwidth]{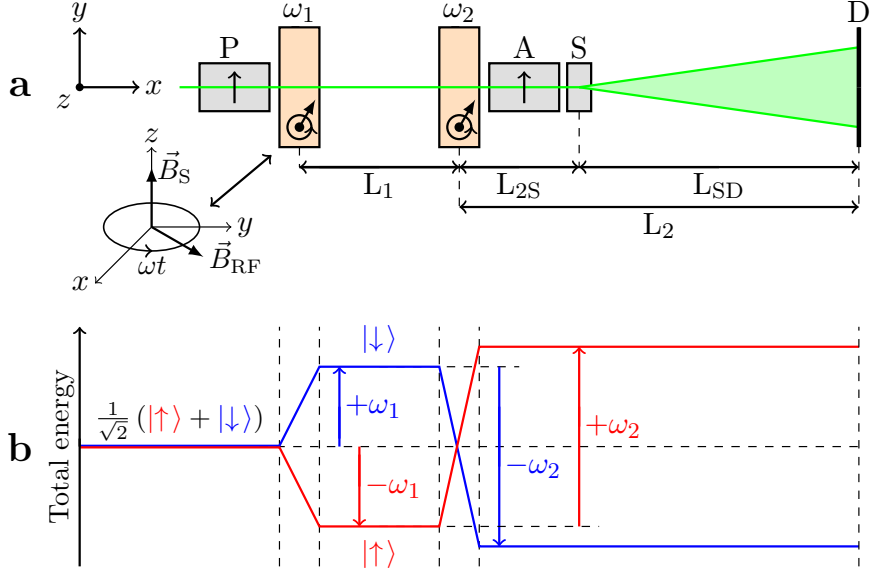}
	\caption{\label{fig:mieze_setup}\textbf{(a)} Sketch of a typical MIEZE setup. Along their flight path (from left to right), neutrons are first spin polarized (P) and manipulated by a sequence of two RFSF operated at field frequencies $\omega_{\rm 1}$ and $\omega_{\rm 2} \neq \omega_{\rm 1}$, respectively. Neutrons' spin are then analyzed (A), before being scattered by a sample (S) and detected by a time-resolved (ToF) detector (D). {\bf Inset:} Superposition of a static field $\vec{B}_{\rm S}$ and radio-frequency field $\vec{B}_{\rm RF}$ as produced by the RFSF. $\vec{B}_{\rm RF}$ is rotating in the plane perpendicular to $\vec{B}_{\rm S}$ at an angular frequency $\omega = \gamma_{\rm n} |\vec{B}_{\rm S}|$, where $\gamma_{\rm n} = 2\pi \cdot 2.916\,$kHz$\cdot$G$^{-1}$ is the neutron gyromagnetic ratio. \textbf{(b)} Energy diagram of a neutron wave packet traveling across the MIEZE setup \cite{Bleuel2005}. The initial wave function is split into two components with opposite spin, as quantified along the static field $z$-direction. This results in an energy difference $\Delta E = 2\hbar\omega_{1}$, which is reversed at the second flipper, yielding $\Delta E = 2\hbar(\omega_{2}-\omega_{1})$. The recombination of the neutron wave packet takes place at the detector where the spin phase is given by Eq. \ref{eq:phiD}.}
\end{figure}

In the late eighties, R. Golub and R. G\"ahler have proposed an alternative design to solenoid-based NSE spectrometers, relying on the use of compact radio-frequency spin flippers (RFSF) and hence termed Neutron Resonance Spin Echo (NRSE) \cite{Golub1987}. NRSE is now available at different instruments throughout the world (MUSES \cite{Longeville2000} at LLB-Saclay and IN22-ZETA \cite{Klimko2003} at ILL-Grenoble in France, RESEDA \cite{Franz2015} and TRISP \cite{Keller2007} at MLZ-Garching, V2-FLEXX \cite{Groitl2015} at BER II-Berlin in Germany, VIN ROSE \cite{Hino2013} at J-PARC/MLF in Japan) and has opened new experimental perspectives by pushing the usual resolution limits of inelastic scattering \cite{Aynajian2008,Chernyshev2012,Lory2017} and diffraction \cite{Pfleiderer2007,Martin2012}. 
As noticed in the early stages of the development of NRSE, series of resonant neutron spin flips can used as building blocks for alternative spectroscopic methods, in analogy with pulse sequences employed in nuclear magnetic resonance (NMR). MIEZE is an elegant application of this idea \cite{Gaehler1992}. A sketch of a typical MIEZE setup is shown in Fig. \ref{fig:mieze_setup}. It consists in a pair of RFSFs separated by a distance $L_{\rm 1}$ and operated at angular frequencies $\omega_{\rm 1}$ and $\omega_{\rm 2} \neq \omega_{\rm 1}$, respectively. At a distance $L_{\rm 2}$ downstream of the second RFSF, where we choose to place a time-resolved detector, the spin phase of a neutron reads \cite{Golub1994}

\begin{equation}
	\label{eq:phiD}
	\varphi_{\rm D} = \omega_{\rm M} \cdot t_{\rm D}+\frac{2}{v} \cdot \left[\omega_{\rm 2}L_{\rm 1} - \frac{\omega_{\rm M}}{2} \cdot \left(L_{\rm 1}+\underbrace{L_{\rm 2S}+L_{\rm SD}}_{L_{\rm 2}}\right)\right] \quad ,
\end{equation}

where $\omega_{\rm M} = 2(\omega_{\rm 2}-\omega_{\rm 1})$ is the modulation (or MIEZE) angular frequency, $t_{\rm D}$ the absolute detection time, $v$ the neutron velocity, $L_{\rm 2S}$ the distance between the second flipper and the sample and $L_{\rm SD}$ the sample-to-detector distance. The velocity-dependent part of Eq. \ref{eq:phiD} is canceled by fulfilling the focusing condition 

\begin{equation}
	\omega_{\rm M} = 2 \omega_{\rm 2} \cdot \frac{L_{\rm 1}}{L_{\rm 1}+L_{\rm 2S}+L_{\rm SD}} \quad ,
	\label{eq:mieze_condition}
\end{equation}

leading to a purely harmonic phase oscillation $\varphi_{\rm D}(t_{\rm D})$ $= \omega_{\rm M} \cdot t_{\rm D}$ at the detector position, even for a coarsely mono\-chromated beam as prepared by a velocity selector. Placing a spin analyzer (A) between the second RFSF and the detector (D) transforms the \emph{phase oscillation} into an \emph{intensity modulation} $I_{\rm D}(t_{\rm D})=I_{\rm 0}/2\cdot\left\{1+\mathcal{C}\cdot\cos[\varphi_{\rm D}(t_{\rm D})]\right\}$, which can be recorded using a time-resolved detector and where $\mathcal{C}$ is the signal contrast. If the neutron beam interacts with a sample (S), located between the analyzer and the detector, the time-dependent intensity will be modified by its dynamics. Indeed, energy transfers $\hbar \omega$ will induce a delay $\Delta t_{\rm D}$ in the neutron propagation time over the distance $L_{\rm SD}$. Averaging this effect over all possible energy transfers yields a finite contrast

\begin{eqnarray}
	\label{eq:CequalsSqtau}
	\mathcal{C} &\propto& \langle \cos\left(\omega_{\rm M} \Delta t_{\rm D}\right)\rangle_{\hbar\omega} < 1\\
	\nonumber
	\text{with} \quad \Delta t_{\rm D} &\equiv& \frac{m_{\rm n} L_{\rm SD}}{h} \cdot \Delta \lambda = \frac{m_{\rm n}^{2}}{2\pi h^{2}} \, L_{\rm SD} \, \lambda^{3} \, \omega \quad ,
\end{eqnarray}

where $m_{\rm n}$ is the neutron mass and $h$ Planck's constant. Note that the definition of $\Delta t_{\rm D}$ in Eq. \ref{eq:CequalsSqtau} is valid in the case of energy transfers which are centered around $\hbar \omega = 0$ and small with respect to the incoming neutron energy (quasi-elastic scattering). Assuming a $\omega$-symmetric scattering function $\mathcal{S}(q,\omega)$, Eq. \ref{eq:CequalsSqtau} is equivalent to \cite{Keller2002}

\begin{eqnarray}
	\mathcal{C} &\propto& \frac{\int \mathcal{S}(q,\omega) \, \cos\left(\omega \tau\right) \, d\omega}{\int \mathcal{S}(q,\omega) \, d\omega} \equiv \frac{\mathcal{S}(q,\tau)}{\mathcal{S}(q,0)}\\  
	\nonumber
	\text{with} \quad \tau &=& \frac{m_{\rm n}^2}{2\pi h^2} \omega_{\rm M} L_{\rm SD} \lambda^3 \quad . 
	\label{eq:FT_sqw}
\end{eqnarray}

In a MIEZE experiment, in full analogy with NSE spectroscopy, we thus have access to the intermediate scattering function, probing sample dynamics over time scales given by the instrumental Fourier time $\tau$, which can reach several 100 ns. In contrast to usual neutron spectroscopy techniques (three-axis, time-of-flight or backscattering), this high-resolution is achieved without drastic loss of intensity since the measured signal does not depend on beam monochromaticity (Eqs. \ref{eq:phiD} and \ref{eq:mieze_condition}). More interestingly, the main advantage of MIEZE is that the measurement is not affected by any depolarizing sample (spin incoherent scatterers \cite{Besenboeck1998}, multi-domain ferromagnets \cite{Kindervater2017}, {\it etc.}) or environment ({\it e.g.} large magnetic fields \cite{Kindervater2015}), as opposed to NSE. 

On the downside, high-resolution can only be reached if no spurious spin phase shift is introduced in the problem, such a ones due to imperfections of the spin manipulation devices or to path length differences throughout the setup. Altogether, this results in an experimental contrast which takes the general form \cite{Brandl2011}  

\begin{equation}
	\mathcal{C}(q,\tau) = \underbrace{\mathcal{R}_{\rm coils}(\tau) \cdot \mathcal{R}_{\rm sample}(q,\tau) \cdot \mathcal{R}_{\rm det}(q,\tau)}_{\mathcal{R}} \cdot \frac{\mathcal{S}(q,\tau)}{\mathcal{S}(q,0)} \quad ,
	\label{eq:Cmeas}
\end{equation}

where the overall reduction factor $\mathcal{R}$ has to be properly quantified in order to correct data and deduce the intrinsic intermediate scattering function.

In practice, an analytical form of $\mathcal{R}_{\rm coils}$ can not be obtained since it involves fine details of the field distribution produced by the RFSF. However, the recent proposal to adopt a \emph{longitudinal} (instead of the standard \emph{transverse}) field geometry for the RFSF \cite{Haussler2003} allows keeping this term close to 1 in the experimental limit. This stems from a self-correction of RFSFs field inhomogeneities and a strongly suppressed field integral variation for divergent flight paths, notably reducing the current needed in correction (Fresnel) coils with respect to the NSE case (by a factor 3 at least) \cite{Krautloher2016}. In what follows, we shall work under the assumption $\mathcal{R}_{\rm coils} = 1$, keeping in mind that its actual value has to be measured before the experiment is performed.

The question of quantifying the reduction factors related to path length inhomogeneities due to sample size and detector thickness - $\mathcal{R}_{\rm sample}$ and $\mathcal{R}_{\rm det}$, respectively - has first been tackled numerically in \cite{Hayashida2008}. In Refs. \cite{Brandl2011,Weber2013}, the problem has been further specified by a combination of analytical calculations and Monte-Carlo simulations. 

%%%%%%%%%%%%%%%%%%%%%%%%%%%%%%%%%%%%%%%%%%%%%%%%%%%%%%%%
%%%%%%%%%%%%%%%%%%%%%%%%%%%%%%%%%%%%%%%%%%%%%%%%%%%%%%%%
%%%%%%%%%%%%%%%%%%%%%%%%%%%%%%%%%%%%%%%%%%%%%%%%%%%%%%%%
%%%%%%%%%%%%%%%%%%%%%%%%%%%%%%%%%%%%%%%%%%%%%%%%%%%%%%%%
\begin{figure*}[!ht]
	\centering
	\includegraphics[width=0.98\textwidth]{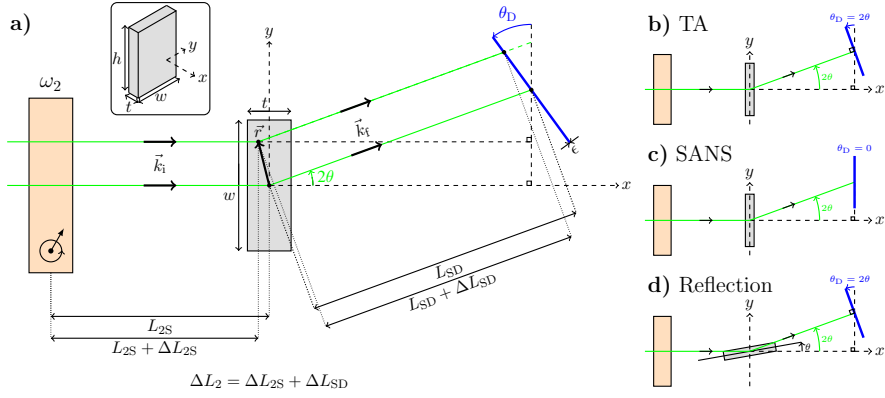}
	\caption{\label{fig:deltaL_MIEZE}\textbf{(a)} Scattering geometry studied in this paper. A plate-like sample of dimensions $w \times t \times h$ (width $\times$ thickness $\times$ height) scatters neutrons at an angle $2\theta$ towards a ToF detector located at a distance $L_{\rm SD}$ downstream. As shown, the main source of path length difference is the sample size, which entails a change in path  $L_{\rm 2S}$ and $L_{\rm SD}$. The various configurations described in the text are shown in \textbf{(b)} (Two-Arms or TA), \textbf{(c)} (Small-Angle or SANS) and \textbf{(d)} (Sample in reflection geometry).}
\end{figure*}
%%%%%%%%%%%%%%%%%%%%%%%%%%%%%%%%%%%%%%%%%%%%%%%%%%%%%%%%
%%%%%%%%%%%%%%%%%%%%%%%%%%%%%%%%%%%%%%%%%%%%%%%%%%%%%%%%
%%%%%%%%%%%%%%%%%%%%%%%%%%%%%%%%%%%%%%%%%%%%%%%%%%%%%%%%
%%%%%%%%%%%%%%%%%%%%%%%%%%%%%%%%%%%%%%%%%%%%%%%%%%%%%%%%

In this paper, we propose more general expressions for $\mathcal{R}_{\rm sample}$ in the case of plate-like samples (parallelepipeds or disks), as usually encountered in soft-matter physics. We treat the case of a parallel incoming beam and the more realistic situation involving finite beam divergence (Sec. \ref{sec:rsample}). In Sec. \ref{sec:rdet}, we calculate the $\mathcal{R}_{\rm det}$ term. Altogether, this allows defining the $(q,\tau)$-range which is accessible under chosen experimental conditions and establishes MIEZE has an interesting counterpart of NSE (Sec. \ref{sec:perfs}). Our findings call for the design of a MIEZE-SANS spectrometer, allowing to study the structure and ps-$\mu$s dynamics of any kind of large scale objects on a single instrument (Sec. \ref{sec:ccl}).   
%%%%%%%%%%%%%%%%%%%%%%%%%%%%%%%%%%%%%%%%%%%%%%%%%%%%%%%%
%%%%%%%%%%%%%%%%%%%%%%%%%%%%%%%%%%%%%%%%%%%%%%%%%%%%%%%%
%%%%%%%%%%%%%%%%%%%%%%%%%%%%%%%%%%%%%%%%%%%%%%%%%%%%%%%%
%%%%%%%%%%%%%%%%%%%%%%%%%%%%%%%%%%%%%%%%%%%%%%%%%%%%%%%%
\section{Path length differences due to sample size and beam divergence}
\label{sec:rsample}

Let us consider a scattering configuration as depicted in Fig. \ref{fig:deltaL_MIEZE}a. In a first step, we shall treat the case of a parallel incoming beam where path length differences with respect to the optical axis read

\begin{equation}
	\Delta L_{\rm 2} = x-\frac{x\cdot\cos\theta_{\rm D}+y\cdot\sin\theta_{\rm D}}{\cos\left(2\theta-\theta_{\rm D}\right)} \quad ,
	\label{eq:pathlengthdiff_sample}
\end{equation}

where $x$ and $y$ are the components of the vector $\vec{r}$ denoting the distance of an arbitrary scattering point to the center of the sample. This leads to a phase difference at the detector given by

\begin{equation}
	\Delta \varphi_{\rm D} = 2\pi \frac{\Delta L_{\rm 2}}{\Lambda} \quad ,
	\label{eq:phasediff_MIEZE}
\end{equation}

where $\Lambda = 2\pi v / \omega_{\rm M}$ is the distance traveled by a neutron of velocity $v$ over one period $2\pi/\omega_{\rm M}$ of the oscillating signal \cite{Brandl2011}. The contrast reduction factor is obtained by averaging $\cos\left(\Delta \varphi_{\rm D}\right)$ over all possible neutron-sample interaction points. We end up with the following reduction factor:

\begin{eqnarray}
	\label{eq:rsample_general}
	\nonumber
	\mathcal{R}_{\rm sample} &\equiv& \frac{\int_{-t/2}^{t/2} \int_{-w/2}^{w/2} \cos \left(\Delta \varphi_{\rm D}\right) dy dx}{w \cdot t}\\
	&=& \sinc\left(\frac{\pi w}{\Lambda}\cdot\frac{\sin \theta_{\rm D}}{\cos(2\theta-\theta_{\rm D})}\right)\\
	\nonumber
	&\times& \sinc\left(\frac{\pi t}{\Lambda}\cdot\left[\frac{\cos\theta_{\rm D}}{\cos(2\theta-\theta_{\rm D})}-1\right]\right) \quad ,
\end{eqnarray}

where we have considered a uniform distribution of $x$ and $y$. In Eq. \ref{eq:rsample_general}, $\sinc$ is the cardinal sine function, $\theta = \arcsin \left(\frac{\lambda q}{4\pi}\right)$ is half the scattering angle for a neutron wavelength $\lambda$ and a momentum transfer $q$, while $\theta_{\rm D}$ is the tilt of the detector measured from the $y$-direction of Fig. \ref{fig:deltaL_MIEZE}. We note that the sample height $h$ does not enter the result, as already explained in Ref. \cite{Brandl2011}.  

Historically, the first MIEZE setups have been installed on two-arms spectrometers \cite{Besenboeck1998,Georgii2011,Haeussler2011,Martin2012a}, for which the ToF detector stays perpendicular to the scattered beam (Fig. \ref{fig:deltaL_MIEZE}b). This is most likely the reason why the authors of Ref. \cite{Brandl2011} have studied the case where $\theta_{\rm D} = 2\theta$ (in what follows, we shall call this configuration 'TA'). Evaluating Eq. \ref{eq:rsample_general} in the TA configuration yields the result previously obtained by Brandl {\it et al.} \cite{Brandl2011}, namely: 

\begin{equation}
	\mathcal{R}_{\rm sample}^{\rm TA} = \sinc\left(\frac{\pi w}{\Lambda}\cdot\sin 2\theta\right) \cdot \sinc\left(\frac{\pi t}{\Lambda}\cdot\left[\cos 2\theta - 1\right]\right) \, .
	\label{eq:rsample_tas}
\end{equation}

However, a quick inspection of Eq. \ref{eq:rsample_general} indicates that setting $\theta_{D} = 0$ will cancel its leading $w$-dependence, {\it i.e.}

\begin{equation}
		\mathcal{R}_{\rm sample}^{\rm SANS} = \sinc \left(\frac{t\pi}{\Lambda} \cdot \left[\frac{1}{\cos 2\theta}-1\right]\right) \, .
		\label{eq:rsample_sans}
\end{equation}

\begin{figure}[!ht]
	\includegraphics[width=0.98\textwidth]{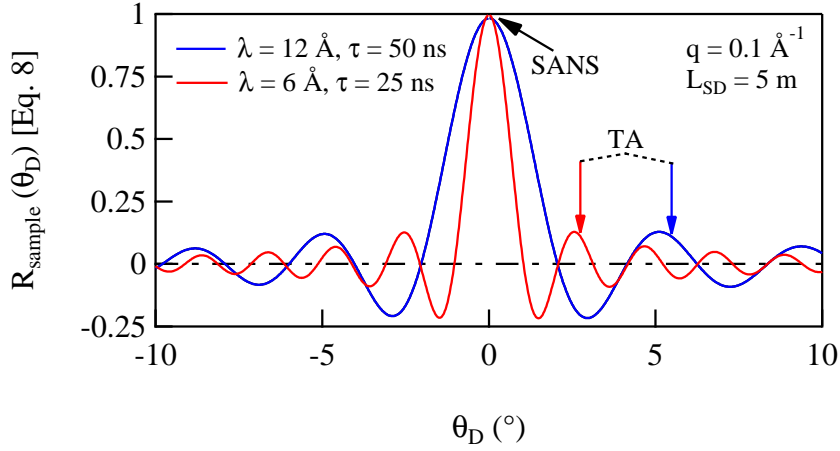}
	\caption{\label{fig:r_vs_thetaD}$\theta_{\rm D}$-dependence of $\mathcal{R}_{\rm sample}$ (Eq. \ref{eq:rsample_general}) for neutron wavelengths $\lambda = 6$ and $12$ \AA~ and Fourier time $\tau = 25$ and $50$ ns, respectively, assuming a momentum transfer $q = 0.1\,$\AA$^{-1}$ and a sample-to-detector distance $L_{\rm SD} = 5$ m. We model a sample of dimensions 10 $\times$ 2  ($w \times t$) mm$^{2}$. The optimal $\mathcal{R}_{\rm sample}$ is achieved for $\theta_{\rm D} = 0$ in both cases. Arrows denote the detector tilt for a sub-optimal TA configuration, pointing the advantage of using the SANS configuration.}
\end{figure}

This results in a much slower decrease of $\mathcal{R}_{\rm sample}$ as a function of $2\theta$ (hence $q$) as compared with Eq. \ref{eq:rsample_tas}, since the effect of the largest sample dimension ({\it i.e.} its width $w$) is canceled. Such configuration shall be referred to as 'SANS', since it mimics a standard small-angle neutron scattering setup for which the position-sensitive detector is perpendicular to the direct beam (Fig. \ref{fig:deltaL_MIEZE}c). We note that since the SANS configuration allows eliminating the effect of all sample dimensions parallel to the detector plane, Eq. \ref{eq:rsample_sans} equally applies to disks or any shapes for which the thickness $t$ is constant. The path length focusing, or in other words the parallelism of the sample flat surface with respect to the detector plane, can be checked by scanning $\theta_{\rm D}$ and optimizing the recorded signal contrast (see Fig. \ref{fig:r_vs_thetaD}).

For the sake of completeness, we mention that Refs. \cite{Besenboeck1998,Keller2002,Brandl2011} have pointed out that path length differences can be also diminished by placing the sample in a reflection geometry, {\it i.e.} rotating it to an angle $\theta$ away from the $x$-direction (Fig. \ref{fig:deltaL_MIEZE}d), yiedling

\begin{equation}
		\mathcal{R}_{\rm sample}^{\rm Reflection} = \sinc \left(\frac{2\pi t \sin \theta}{\Lambda}\right) \, .
		\label{eq:rsample_refl}
\end{equation} 

However, such configuration is likely to be unpractical, for instance when samples are placed on a sample changer. Moreover, it entails a loss of usable flux (proportional to $\sin \theta$) such that the SANS configuration should be preferred in most cases. 

For clarity, we postpone to \ref{sec:appendix} the discussion addressing cases where the sample can neither be positioned with its faces perpendicular to the incoming beam, nor in reflection geometry. It leads to an expression which allows recovering Eqs. \ref{eq:rsample_tas}. \ref{eq:rsample_sans} and \ref{eq:rsample_refl} and finding the optimal detector tilt angle in any conceivable configuration.

Turning back to the angular dependence of Eq. \ref{eq:rsample_general}, we notice that it can be exploited to estimate the maximum in-plane beam divergence which can be experimentally tolerated. This is done by expanding the argument of the $w$-term in Eq. \ref{eq:rsample_general} to second order and averaging the result over the distribution of angular offsets, modeled by a Gaussian function of full-width at half-maximum $\beta$:

\begin{equation}
	\mathcal{R}_{\rm sample}^{\rm SANS}(\beta) \sim \mathcal{R}_{\rm sample}^{\rm SANS}(0) \cdot \left(1 - \frac{\pi^2}{12 \ln 16}\cdot\frac{w^{2} \beta^{2}}{\Lambda^{2} \cos^{2} 2\theta}\right) \quad .
	\label{eq:rsample_beta}
\end{equation}

In a typical SANS measurement, the collimation length $L_{\rm C}$ should match the sample-to-detector distance $L_{\rm SD}$. For slit sizes equal to the sample width $w$, $\beta \sim w / L_{\rm SD}$ and we get a divergence-induced reduction of $\mathcal{R}_{\rm sample}$ of less than 1 \% for the parameters used in Fig. \ref{fig:r_vs_thetaD}. Taking $\lambda = 12$ \AA~and $\tau = 250$ ns, one still gets $\mathcal{R}_{\rm sample}(\beta)/\mathcal{R}_{\rm sample}(0)$ of the order of 3 \%. This clearly leaves space for increasing the incoming beam divergence (and thus usable neutron flux) while maintaining high time-resolution, recalling that the associated spin phase inhomogeneities could also be well-compensated using a longitudinal field geometry with correction (Fresnel) coils \cite{Krautloher2016}. 
%%%%%%%%%%%%%%%%%%%%%%%%%%%%%%%%%%%%%%%%%%%%%%%%%%%%%%%%
%%%%%%%%%%%%%%%%%%%%%%%%%%%%%%%%%%%%%%%%%%%%%%%%%%%%%%%%
%%%%%%%%%%%%%%%%%%%%%%%%%%%%%%%%%%%%%%%%%%%%%%%%%%%%%%%%
%%%%%%%%%%%%%%%%%%%%%%%%%%%%%%%%%%%%%%%%%%%%%%%%%%%%%%%%
\section{Path length differences due to detector thickness}
\label{sec:rdet}

Another unavoidable source of path length difference is the finite thickness $\epsilon$ of the detector. In practice, it can be made small by using thin $^{\rm 10}$B conversion layers but should remain in the 10 $\mu$m-range for achieving decent efficiencies. The corresponding reduction factor is given by
.
\begin{equation}
		\mathcal{R}_{\rm det} = \sinc \left(\frac{\epsilon \pi}{\Lambda \cdot \cos\left[2\theta-\theta_{\rm D}\right]}\right) \quad ,
		\label{eq:rdet_all}
\end{equation}

under the assumption of a constant detection probability across the detector thickness. In the TA configuration ($\theta_{\rm D} = 2\theta$), $\mathcal{R}_{\rm det}$ is constant while it will decrease upon scattering angle (or $q$) increase for the SANS case. However, the $\cos$-term renders this effect negligible in the small-angle limit.
%%%%%%%%%%%%%%%%%%%%%%%%%%%%%%%%%%%%%%%%%%%%%%%%%%%%%%%%
%%%%%%%%%%%%%%%%%%%%%%%%%%%%%%%%%%%%%%%%%%%%%%%%%%%%%%%%
%%%%%%%%%%%%%%%%%%%%%%%%%%%%%%%%%%%%%%%%%%%%%%%%%%%%%%%%
%%%%%%%%%%%%%%%%%%%%%%%%%%%%%%%%%%%%%%%%%%%%%%%%%%%%%%%%
\begin{figure*}[t]
	\centering
	\includegraphics[width=0.98\textwidth]{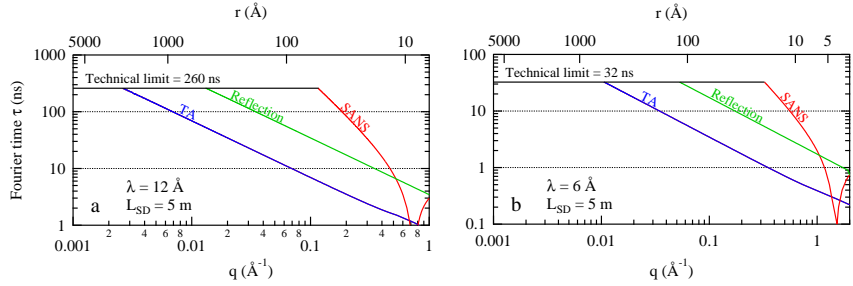}
	\caption{\label{fig:tau_vs_q}Accessible $(q,\tau)$-range \textbf{(a)} for $\lambda = 12$\, \AA \, and \textbf{(b)} for $\lambda = 6\,$\AA, with a sample-to-detector distance $L_{\rm SD} = 5\,$m, a detector thickness $\epsilon = 10\,\mu$m and a sample of dimensions 10 $\times$ 2 ($w \times t$) mm$^2$. It is seen that the SANS configuration offers an access to the largest parameter space as compared with TAS and Reflection geometry. The dip in the SANS curves stem from the $1/\cos 2\theta$-dependence of Eq. \ref{eq:rdet_all}. Only in that $q$-range, the Reflection geometry should be preferred.}
\end{figure*}
%%%%%%%%%%%%%%%%%%%%%%%%%%%%%%%%%%%%%%%%%%%%%%%%%%%%%%%%
%%%%%%%%%%%%%%%%%%%%%%%%%%%%%%%%%%%%%%%%%%%%%%%%%%%%%%%%
%%%%%%%%%%%%%%%%%%%%%%%%%%%%%%%%%%%%%%%%%%%%%%%%%%%%%%%%
%%%%%%%%%%%%%%%%%%%%%%%%%%%%%%%%%%%%%%%%%%%%%%%%%%%%%%%%
\section{Expected performance of a MIEZE-SANS setup}
\label{sec:perfs}

In order to estimate the accessible $(q,\tau)$-range for a quasi-elastic experiment using a MIEZE spectrometer, we calculate the product $\mathcal{R} = \mathcal{R}_{\rm coils} \cdot \mathcal{R}_{\rm sample} \cdot \mathcal{R}_{\rm det}$ (Eq. \ref{eq:Cmeas}), with $\mathcal{R}_{\rm coils} = 1$, for series of momentum transfers $q$ as a function of Fourier time $\tau$ in the TA, SANS and Reflection configurations. This allows defining accessible parameter ranges by tracing the $\mathcal{R} = 1/3$ line, a commonly employed low-limit in NSE spectroscopy. In Fig. \ref{fig:tau_vs_q}, we give the outcome of such procedure for two representative neutron wavelength ($\lambda = 6$ and $12$ \AA) and a sample-to-detector distance $L_{\rm SD} = 5$ m. The technical upper limit for $\tau$ is calculated using Eq. \ref{eq:FT_sqw}, assuming a maximum RF field angular frequency $\omega_{\rm 2} = 2\pi \cdot 10\,\text{MHz}$ with $L_{\rm 1}$ = 2 m and $L_{\rm 2S}$ = 1.5 m as inputs of Eq. \ref{eq:mieze_condition}. In a large $q$-range, we clearly see that the SANS configuration gives access to much better time-resolution than the TA configuration. Only for momentum transfers corresponding to scattering angles close to $\pi / 2$, the reflection configuration is superior to the SANS one.

\begin{figure}[!ht]
	\centering
	\includegraphics[width=0.98\textwidth]{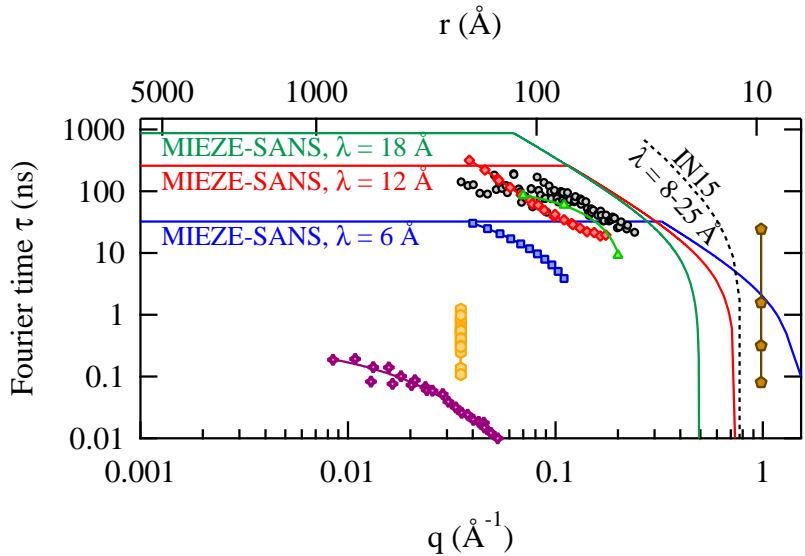}
	\caption{\label{fig:tau_vs_q_examples}Accessible $(q,\tau)$-range with MIEZE-SANS at various neutron wavelengths, for a sample-to-detector distance $L_{\rm SD} = 5\,$m, a detector thickness $\epsilon = 10 \mu$m and a sample of dimension 10 $\times$ 2 ($w \times t$) mm$^2$ (solid lines). The overlayed symbols represent a series of NSE results collected from the recent literature.	On the large time side, we show typical time-scales of processes involved in the dynamics of protein domains (red diamonds from \cite{Callaway2013} ), grafted-polymers (green triangles from \cite{Mark2017} and blue squares from \cite{Poling2017}) and hemoglobin (black circles from \cite{Longeville2017}). On the short time side, more typical of magnetic systems, we provide examples of spin fluctuation linewidths as observed at the Curie point in Fe (pink crosses from \cite{Kindervater2017}), as a function of magnetic field in the chiral magnet MnSi (yellow hexagons from \cite{Pappas2017}) and due to soliton dynamics in antiferromagnetic spin chains (brown pentagons from \cite{Boucher1985}). For the sake of comparison, we also plot the upper $(q,\tau)$ limits of the reference NSE spectrometer IN15 (Institut Laue Langevin, France) operated at wavelengths in the 8-25 \AA\,range (black dashed curve from \cite{Schleger1999}).}
\end{figure}

For an objective estimation the viability of the MIEZE technique, it is important to look for experimental situations which require the most extreme conditions in terms of momentum transfer and time-resolution. Of all fields using NSE as a paramount experimental technique, macro\-molecular physics is pro\-bably the most de\-manding in terms of accessible $(q,\tau)$-range. This stems from the fact that the studied objects are usually nano\-sized, while showing dy\-namics up to the high 100 ns range. We have collected examples from recent literature on these topics, dealing with protein domain motions \cite{Callaway2013}, polymer-grafted nano-particles dynamics \cite{Mark2017,Poling2017} and hemoglobin diffusion \cite{Longeville2017}. In order to check the feasibility of quasi-elastic scattering experiments using the MIEZE method, we compare the characteristic times $\tau_{\rm 0} = 1/(D q^2)$ (where $D$ is the mea\-sured diffusion coefficient) obtained by classical NSE with the experimental range covered by MIEZE in a SANS configuration. As seen on Fig. \ref{fig:tau_vs_q_examples}, it is clear that MIEZE can be considered as a fair competitor to classical NSE in terms of dynamic range. Moreover, we recall that a MIEZE measurement is not affected by spin incoherent scattering, thus avoiding the recourse to skilful deu\-teration schemes. In the case of fully protonated samples and for the same neutron intensity $I_{\rm 0}$, MIEZE would offer a gain in efficiency of a factor 9 with respect to NSE considering the standard figure of merit $I_{\rm 0} \cdot \mathcal{C}^{2}$. In practice, this number has to be slightly reduced since time-resolved detectors usually have smaller efficiencies than $^{3}$He tubes commonly used on NSE spectrometers.

In Fig. \ref{fig:tau_vs_q_examples}, we also report on examples taken from the field of magnetism. Despite the relatively fast characteristic times, standard NSE experiments are usually precluded on multi-domain ferromagnets \cite{Kindervater2017} or under magnetic fields \cite{Boucher1985,Pappas2017}, and the literature remains scarce on the topic. Since MIEZE will not suffer from such experimental conditions, we anticipate that it will play an important role in the field of magnetism in a near future, for instance in studies of field-induced quantum criticality and of excitations emerging from topological defects in ferro- or helimagnets. 

%%%%%%%%%%%%%%%%%%%%%%%%%%%%%%%%%%%%%%%%%%%%%%%%%%%%%%%%
%%%%%%%%%%%%%%%%%%%%%%%%%%%%%%%%%%%%%%%%%%%%%%%%%%%%%%%%
%%%%%%%%%%%%%%%%%%%%%%%%%%%%%%%%%%%%%%%%%%%%%%%%%%%%%%%%
%%%%%%%%%%%%%%%%%%%%%%%%%%%%%%%%%%%%%%%%%%%%%%%%%%%%%%%%
\section{Conclusions}
\label{sec:ccl}

We have provided an analytical framework which allows determining the actual trade-off between intensity and time-resolution of a MIEZE spectrometer. By means of simple geometrical arguments, we have shown that, in a SANS geometry, the resolution of MIEZE is fairly competitive with respect to traditional NSE. Since its performance does not suffer from depolarization by the sample or its environment, the method thus has the potential to enlarge the field of high-resolution neutron spectroscopy. We conclude that the implementation of MIEZE on a traditional polarized SANS instrument could contribute in addressing some of the yet unexplored questions in modern soft- and hard-condensed matter physics. On existing two-axis instruments, the time-resolution would also be drastically improved by \emph{simply} ins\-talling the time-resolved detector on a rotation stage.  
%%%%%%%%%%%%%%%%%%%%%%%%%%%%%%%%%%%%%%%%%%%%%%%%%%%%%%%%
%%%%%%%%%%%%%%%%%%%%%%%%%%%%%%%%%%%%%%%%%%%%%%%%%%%%%%%%
%%%%%%%%%%%%%%%%%%%%%%%%%%%%%%%%%%%%%%%%%%%%%%%%%%%%%%%%
%%%%%%%%%%%%%%%%%%%%%%%%%%%%%%%%%%%%%%%%%%%%%%%%%%%%%%%%
\section{Acknowledgements}
I wish to thank Annie Br\^u\-let, Gr\'egory Chaboussant, Jacques Jestin, St\'ephane Lon\-geville and Fr\'ed\'eric Ott for several discussions which have incidentally triggered this work and their help during the preparation of the present manuscript. The continuous support from Christiane Alba-Simio\-nesco, Isabelle Mirebeau and Emmy Th\'evenot Martin is also much appre\-ciated. 
%%%%%%%%%%%%%%%%%%%%%%%%%%%%%%%%%%%%%%%%%%%%%%%%%%%%%%%%
%%%%%%%%%%%%%%%%%%%%%%%%%%%%%%%%%%%%%%%%%%%%%%%%%%%%%%%%
%%%%%%%%%%%%%%%%%%%%%%%%%%%%%%%%%%%%%%%%%%%%%%%%%%%%%%%%
%%%%%%%%%%%%%%%%%%%%%%%%%%%%%%%%%%%%%%%%%%%%%%%%%%%%%%%%
\renewcommand\thefigure{\thesection.\arabic{figure}} 
\appendix
\setcounter{figure}{0} 
\section{General expression for the sample reduction factor}
\label{sec:appendix}

\begin{figure}[!ht]
	\centering
	\includegraphics[width=0.98\textwidth]{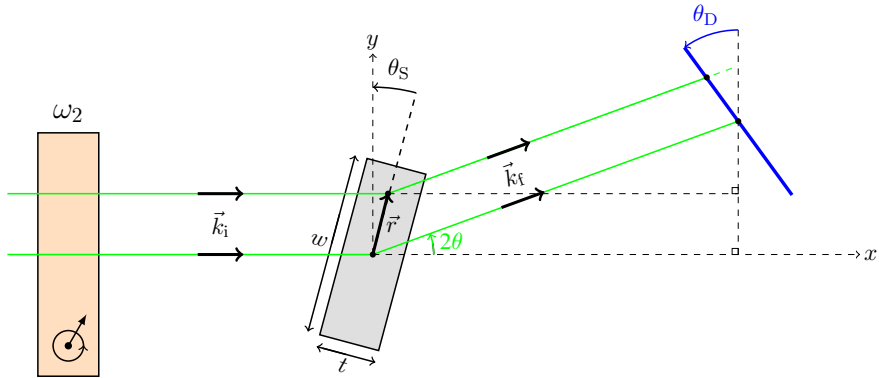}
	\caption{\label{fig:pathlengthsamplethetaS}Same scattering geometry as considered in main text (Fig. \ref{fig:deltaL_MIEZE}), where we allow for a sample tilt by an angle $\theta_{\rm S}$ with respect to the $y$-direction.}
\end{figure}

Under certain circumstances, for instance when working with oriented samples \cite{Marry2013}, it might be impossible to align the sample surface perpendicular to the incoming beam or in reflection geometry. The corresponding expression for the sample reduction factor is obtained from Eq. \ref{eq:pathlengthdiff_sample} by means of an appropriate change of variables

\begin{eqnarray}
	\nonumber
	\mathcal{R}_{\rm sample} &=& \sinc\left(\frac{\pi t}{\Lambda}\cdot\left[\cos\theta_{\rm S}-\frac{\cos\left(\theta_{\rm D}-\theta_{\rm S}\right)}{\cos\left(2\theta-\theta_{\rm D}\right)}\right]\right)\\
	\nonumber
	&\times& \sinc\left(\frac{\pi w}{\Lambda}\cdot\left[\sin\theta_{\rm S}+\frac{\sin\left(\theta_{\rm D}-\theta_{\rm S}\right)}{\cos\left(2\theta-\theta_{\rm D}\right)}\right]\right) \quad ,\\
	\label{eq:rsample_thetaS}
\end{eqnarray}		

where symbols bear the same meaning as in main text and $\theta_{\rm S}$ is the sample tilt angle with respect to the $y$-direction (see Fig. \ref{fig:pathlengthsamplethetaS}). 

%Setting $\left\{\theta_{\rm D} = 2\theta, \theta_{\rm S} = 0\right\}$, $\left\{\theta_{\rm D} = \theta_{\rm S} = 0\right\}$ or $\left\{\theta_{\rm D} = 2\theta, \theta_{\rm S} = \pi/2 + \theta\right\}$ allows recovering the expressions given in main text for the TA (Eq. \ref{eq:rsample_tas}), SANS (Eq. \ref{eq:rsample_sans}) and Reflection (Eq. \ref{eq:rsample_refl}) configuration, respectively

If we seek for the value of $\theta_{\rm D}$ which maximizes $\mathcal{R}_{\rm sample}$, meaning that we try to cancel the arguments of the $\sinc$ functions in Eq. \ref{eq:rsample_thetaS}, we get

\begin{equation}
	\theta_{\rm D} = \arctan\left(\frac{1-\cos 2\theta}{\sin 2\theta-\tan\theta_{\rm S}}\right)
	\label{eq:thetaDt}
\end{equation} 

and

\begin{equation}
	\theta_{\rm D} = \arctan\left(\frac{\tan\theta_{\rm S}\cdot\left[1-\cos 2\theta\right]}{\tan\theta_{\rm S}\cdot\sin 2\theta+1}\right)
	\label{eq:thetaDw}
\end{equation} 
		
for the first and second member, respectively. Clearly, Eqs. \ref{eq:thetaDt} and \ref{eq:thetaDw} can not be simultaneously satisfied. The optimal experimental strategy thus consists in fulfilling Eq. \ref{eq:thetaDw} in order to suppress the strongest effect, due to the sample's width $w$ assumed to be much larger than its thickness $t$.

%%%%%%%%%%%%%%%%%%%%%%%%%%%%%%%%%%%%%%%%%%%%%%%%%%%%%%%%
%%%%%%%%%%%%%%%%%%%%%%%%%%%%%%%%%%%%%%%%%%%%%%%%%%%%%%%%
%%%%%%%%%%%%%%%%%%%%%%%%%%%%%%%%%%%%%%%%%%%%%%%%%%%%%%%%
%%%%%%%%%%%%%%%%%%%%%%%%%%%%%%%%%%%%%%%%%%%%%%%%%%%%%%%%
%\clearpage
\bibliographystyle{elsarticle-num}
%\bibliography{MIEZE_resolution}

\begin{thebibliography}{10}
\expandafter\ifx\csname url\endcsname\relax
  \def\url#1{\texttt{#1}}\fi
\expandafter\ifx\csname urlprefix\endcsname\relax\def\urlprefix{URL }\fi
\expandafter\ifx\csname href\endcsname\relax
  \def\href#1#2{#2} \def\path#1{#1}\fi

\bibitem{Schleger1999}
P.~Schleger, G.~Ehlers, A.~Kollmar, B.~Alefeld, J.~Barthelemy, H.~Casalta,
  B.~Farago, P.~Giraud, C.~Hayes, C.~Lartigue, F.~Mezei, D.~Richter, The
  sub-ne{V} resolution {NSE} spectrometer {IN15} at the {I}nstitute {L}aue
  {L}angevin, Physica B: Condensed Matter 266~(1) (1999) 49 -- 55.
\newblock \href {http://dx.doi.org/10.1016/S0921-4526(98)01491-4}
  {\path{doi:10.1016/S0921-4526(98)01491-4}}.

\bibitem{Besenboeck1998}
W.~Besenb\"ock, R.~G\"ahler, P.~Hank, R.~Kahn, M.~K\"oppe, C.-H.~D. Novion,
  W.~Petry, J.~Wuttke, First scattering experiment on {MIEZE}: {A} {F}ourier
  transform time-of-flight spectrometer using resonance coils, Journal of
  Neutron Research 7~(1) (1998) 65--74.
\newblock \href {http://dx.doi.org/10.1080/10238169808200231}
  {\path{doi:10.1080/10238169808200231}}.

\bibitem{Kindervater2017}
J.~Kindervater, S.~S\"aubert, P.~B\"oni, Dipolar effects on the critical
  fluctuations in {F}e: Investigation by the neutron spin-echo technique
  {MIEZE}, Phys. Rev. B 95 (2017) 014429.
\newblock \href {http://dx.doi.org/10.1103/PhysRevB.95.014429}
  {\path{doi:10.1103/PhysRevB.95.014429}}.

\bibitem{Kindervater2015}
{Kindervater, J.}, {Martin, N.}, {H\"aussler, W.}, {Krautloher, M.}, {Fuchs,
  C.}, {M\"uhlbauer, S.}, {Lim, J.A.}, {Blackburn, E.}, {B\"oni, P.},
  {Pfleiderer, C.}, Neutron spin echo spectroscopy under 17 {T} magnetic field
  at {RESEDA}, EPJ Web of Conferences 83 (2015) 03008.
\newblock \href {http://dx.doi.org/10.1051/epjconf/20158303008}
  {\path{doi:10.1051/epjconf/20158303008}}.

\bibitem{Bleuel2005}
M.~Bleuel, K.~Littrell, R.~G\"ahler, J.~Lal, {MISANS}, a method for
  quasi-elastic small angle neutron scattering experiments, Physica B: Cond.
  Matt. 356~(1-4) (2005) 213--217.

\bibitem{Golub1987}
R.~Golub, R.~G\"ahler, A {N}eutron {R}esonance {S}pin {E}cho {S}pectrometer for
  {Q}uasi-{E}lastic and {I}nelastic {S}cattering, Phys. Lett. A 123~(1) (1987)
  43--48.
\newblock \href {http://dx.doi.org/10.1016/0375-9601(87)90760-2}
  {\path{doi:10.1016/0375-9601(87)90760-2}}.

\bibitem{Longeville2000}
S.~Longeville, La spectroscopie neutronique \`a \'echo de spin \`a champ nul ou
  par r\'esonance, J. Phys. IV France 10 (2000) 59--75.
\newblock \href {http://dx.doi.org/10.1051/jp4:2000105}
  {\path{doi:10.1051/jp4:2000105}}.

\bibitem{Klimko2003}
S.~Klimko, C.~Stadler, P.~B\"oni, R.~Currat, F.~Demmel, B.~F\r{a}k,
  R.~G\"ahler, F.~Mezei, B.~Toperverg, Implementation of a zero-field spin-echo
  option at the three-axis spectrometer {IN3} ({ILL}, {G}renoble) and first
  application for measurements of phonon line widths in superfluid $^{4}${H}e,
  Physica B: Condensed Matter 335~(1) (2003) 188 -- 192, proceedings of the
  Fourth International Workshop on Polarised Neutrons for Condensed Matter
  Investigations.
\newblock \href
  {http://dx.doi.org/https://doi.org/10.1016/S0921-4526(03)00234-5}
  {\path{doi:https://doi.org/10.1016/S0921-4526(03)00234-5}}.

\bibitem{Franz2015}
C.~Franz, T.~Schr\"oder, Reseda: Resonance spin echo spectrometer, Journal of
  large-scale research facilities\href {http://dx.doi.org/10.17815/jlsrf-1-37}
  {\path{doi:10.17815/jlsrf-1-37}}.

\bibitem{Keller2007}
T.~Keller, P.~Aynajian, S.~Bayrakci, K.~Buchner, K.~Habicht, H.~Klann, M.~Ohl,
  B.~Keimer, Scientific {R}eview: {T}he {T}riple {A}xis {S}pin-{E}cho
  {S}pectrometer {TRISP} at the {FRM} {II}, Neutron News 18~(2) (2007) 16--18.
\newblock \href {http://dx.doi.org/10.1080/10448630701328372}
  {\path{doi:10.1080/10448630701328372}}.

\bibitem{Groitl2015}
F.~Groitl, T.~Keller, D.~L. Quintero-Castro, K.~Habicht, Neutron resonance
  spin-echo upgrade at the three-axis spectrometer {FLEXX}, Review of
  Scientific Instruments 86~(2) (2015) 025110.
\newblock \href {http://dx.doi.org/10.1063/1.4908167}
  {\path{doi:10.1063/1.4908167}}.

\bibitem{Hino2013}
M.~Hino, T.~Oda, M.~Kitaguchi, N.~L. Yamada, H.~Sagehashi, Y.~Kawabata,
  H.~Seto, Current {S}tatus of {BL}06 {B}eam {L}ine for {VIN} {ROSE} at
  {J}-{PARC}/{MLF}, Physics Procedia 42 (2013) 136 -- 141, 9th International
  Conference on Polarised Neutrons in Condensed Matter Investigations.
\newblock \href {http://dx.doi.org/10.1016/j.phpro.2013.03.187}
  {\path{doi:10.1016/j.phpro.2013.03.187}}.

\bibitem{Aynajian2008}
P.~Aynajian, T.~Keller, L.~Boeri, S.~M. Shapiro, K.~Habicht, B.~Keimer, Energy
  {G}aps and {K}ohn {A}nomalies in {E}lemental {S}uperconductors, Science
  319~(5869) (2008) 1509--1512.
\newblock \href {http://dx.doi.org/10.1126/science.1154115}
  {\path{doi:10.1126/science.1154115}}.

\bibitem{Chernyshev2012}
A.~L. Chernyshev, M.~E. Zhitomirsky, N.~Martin, L.-P. Regnault, {L}ifetime of
  {G}apped {E}xcitations in a {C}ollinear {Q}uantum {A}ntiferromagnet, Phys.
  Rev. Lett. 109 (2012) 097201.
\newblock \href {http://dx.doi.org/10.1103/PhysRevLett.109.097201}
  {\path{doi:10.1103/PhysRevLett.109.097201}}.

\bibitem{Lory2017}
P.-F. Lory, S.~Pailh\`es, V.~M. Giordano, H.~Euchner, H.~D. Nguyen, R.~Ramlau,
  H.~Borrmann, M.~Schmidt, M.~Baitinger, M.~Ikeda, P.~Tomec, M.~Mihalkovic,
  C.~Allio, M.~R. Johnson, H.~Schober, Y.~Sidis, F.~Bourdarot, L.~P. Regnault,
  J.~Ollivier, S.~Paschen, Y.~Grin, M.~de~Boissieu, Direct measurement of
  individual phonon lifetimes in the clathrate compound
  {B}a$_{7.81}${G}e$_{40.67}${A}u$_{5.33}$, Nature Communications 8~(1) (2017)
  491.
\newblock \href {http://dx.doi.org/10.1038/s41467-017-00584-7}
  {\path{doi:10.1038/s41467-017-00584-7}}.

\bibitem{Pfleiderer2007}
C.~Pfleiderer, P.~B{\"o}ni, T.~Keller, U.~K. R{\"o}{\ss}ler, A.~Rosch,
  Non-{F}ermi {L}iquid {M}etal {W}ithout {Q}uantum {C}riticality, Science
  316~(5833) (2007) 1871--1874.
\newblock \href {http://dx.doi.org/10.1126/science.1142644}
  {\path{doi:10.1126/science.1142644}}.

\bibitem{Martin2012}
N.~Martin, L.-P. Regnault, S.~Klimko, Neutron {L}armor {D}iffraction study of
  the {B}a{M}$_{2}$({XO}$_4$)$_{2}$ ({M} = {C}o, {N}i; {X} = {A}s, {P})
  compounds, Journal of Physics: Conference Series 340~(1) (2012) 012012.
\newblock \href {http://dx.doi.org/10.1088/1742-6596/340/1/012012}
  {\path{doi:10.1088/1742-6596/340/1/012012}}.

\bibitem{Gaehler1992}
R.~G\"ahler, R.~Golub, T.~Keller, Neutron resonance spin echo$-$a new tool for
  high resolution spectroscopy, Physica B: Condensed Matter 180 (1992) 899 --
  902.
\newblock \href {http://dx.doi.org/10.1016/0921-4526(92)90503-K}
  {\path{doi:10.1016/0921-4526(92)90503-K}}.

\bibitem{Golub1994}
R.~Golub, R.~G\"ahler, T.~Keller, A plane wave approach to particle beam
  magnetic resonance, American Journal of Physics 62~(9) (1994) 779--788.
\newblock \href {http://dx.doi.org/10.1119/1.17459}
  {\path{doi:10.1119/1.17459}}.

\bibitem{Keller2002}
T.~Keller, R.~Golub, R.~G\"ahler, Neutron {S}pin {E}cho$-${A} {T}echnique for
  {H}igh-{R}esolution {N}eutron {S}cattering, in: R.~Pike, P.~Sabatier (Eds.),
  Scattering, Academic Press, London, 2002, pp. 1264 -- 1286.
\newblock \href {http://dx.doi.org/10.1016/B978-012613760-6/50068-1}
  {\path{doi:10.1016/B978-012613760-6/50068-1}}.

\bibitem{Brandl2011}
G.~Brandl, R.~Georgii, W.~H\"au\ss{}ler, S.~M\"uhlbauer, P.~B\"oni, Large
  scales-long times: Adding high energy resolution to {SANS}, Nuclear
  Instruments and Methods in Physics Research Section A: Accelerators,
  Spectrometers, Detectors and Associated Equipment 654~(1) (2011) 394 -- 398.
\newblock \href {http://dx.doi.org/10.1016/j.nima.2011.07.003}
  {\path{doi:10.1016/j.nima.2011.07.003}}.

\bibitem{Haussler2003}
W.~H\"aussler, U.~Schmidt, G.~Ehlers, F.~Mezei, Neutron resonance spin echo
  using spin echo correction coils, Chem. Phys. 292 (2003) 501--510.
\newblock \href {http://dx.doi.org/10.1016/S0301-0104(03)00119-8}
  {\path{doi:10.1016/S0301-0104(03)00119-8}}.

\bibitem{Krautloher2016}
M.~Krautloher, J.~Kindervater, T.~Keller, W.~H\"aussler, Neutron resonance spin
  echo with longitudinal {DC} fields, Review of Scientific Instruments 87~(12)
  (2016) 125110.
\newblock \href {http://dx.doi.org/10.1063/1.4972395}
  {\path{doi:10.1063/1.4972395}}.

\bibitem{Hayashida2008}
H.~Hayashida, M.~Hino, M.~Kitaguchi, Y.~Kawabata, N.~Achiwa, A study of
  resolution function on a {MIEZE} spectrometer, Measurement Science and
  Technology 19~(3) (2008) 034006.

\bibitem{Weber2013}
T.~Weber, G.~Brandl, R.~Georgii, W.~H\"au\ss{}ler, S.~Weichselbaumer,
  P.~B\"oni, Monte-{C}arlo simulations for the optimisation of a {TOF}-{MIEZE}
  instrument, Nuclear Instruments and Methods in Physics Research Section A:
  Accelerators, Spectrometers, Detectors and Associated Equipment 713 (2013)
  71--75.
\newblock \href {http://dx.doi.org/10.1016/j.nima.2013.03.010}
  {\path{doi:10.1016/j.nima.2013.03.010}}.

\bibitem{Georgii2011}
R.~Georgii, G.~Brandl, N.~Arend, W.~H\"au\ss{}ler, A.~Tischendorf,
  C.~Pfleiderer, P.~B\"oni, J.~Lal, Turn-key module for neutron scattering with
  sub-$\mu$ev resolution, Applied Physics Letters 98~(7) (2011) 073505.
\newblock \href {http://dx.doi.org/10.1063/1.3556558}
  {\path{doi:10.1063/1.3556558}}.

\bibitem{Haeussler2011}
W.~H\"aussler, P.~B\"oni, M.~Klein, C.~J. Schmidt, U.~Schmidt, F.~Groitl,
  J.~Kindervater, Detection of high frequency intensity oscillations at
  {RESEDA} using the {CASCADE} detector, Review of Scientific Instruments
  82~(4) (2011) 045101.
\newblock \href {http://dx.doi.org/10.1063/1.3571300}
  {\path{doi:10.1063/1.3571300}}.

\bibitem{Martin2012a}
N.~Martin, \href{http://www.theses.fr/2012GRENY116}{{\'E}tude {S}tructurale et
  {D}ynamique de {P}lusieurs {S}yst\`emes {M}agn\'etiques par la {T}echnique de
  l'{\'e}cho de {S}pin {N}eutronique {R}\'esonant}, Ph.D. thesis, Universit\'e
  Joseph Fourier (May 2012).
\newline\urlprefix\url{http://www.theses.fr/2012GRENY116}

\bibitem{Callaway2013}
D.~J.~E. Callaway, B.~Farago, Z.~Bu, Nanoscale protein dynamics: {A} new
  frontier for neutron spin echo spectroscopy, The European Physical Journal E
  36~(7) (2013) 76.
\newblock \href {http://dx.doi.org/10.1140/epje/i2013-13076-1}
  {\path{doi:10.1140/epje/i2013-13076-1}}.

\bibitem{Mark2017}
C.~Mark, O.~Holderer, J.~Allgaier, E.~H\"ubner, W.~Pyckhout-Hintzen,
  M.~Zamponi, A.~Radulescu, A.~Feoktystov, M.~Monkenbusch, N.~Jalarvo,
  D.~Richter, Polymer {C}hain {C}onformation and {D}ynamical {C}onfinement in a
  {M}odel {O}ne-{C}omponent {N}anocomposite, Phys. Rev. Lett. 119 (2017)
  047801.
\newblock \href {http://dx.doi.org/10.1103/PhysRevLett.119.047801}
  {\path{doi:10.1103/PhysRevLett.119.047801}}.

\bibitem{Poling2017}
R.~Poling-Skutvik, K.~N. Olafson, S.~Narayanan, L.~Stingaciu, A.~Faraone, J.~C.
  Conrad, R.~Krishnamoorti, Confined dynamics of grafted polymer chains in
  solutions of linear polymer, Macromolecules 50~(18) (2017) 7372--7379.
\newblock \href {http://dx.doi.org/10.1021/acs.macromol.7b01245}
  {\path{doi:10.1021/acs.macromol.7b01245}}.

\bibitem{Longeville2017}
S.~Longeville, L.-R. Stingaciu, Hemoglobin diffusion and the dynamics of oxygen
  capture by red blood cells, Scientific Reports 7~(1) (2017) 10448.
\newblock \href {http://dx.doi.org/10.1038/s41598-017-09146-9}
  {\path{doi:10.1038/s41598-017-09146-9}}.

\bibitem{Boucher1985}
J.~P. Boucher, F.~Mezei, L.~P. Regnault, J.~P. Renard, Diffusion of {S}olitons
  in the {A}ntiferromagnetic {C}hains of
  ${(\mathrm{C}{\mathrm{D}}_{3})}_{4}${NM}n{C}l$_{3}$: {A} {S}tudy by {N}eutron
  {S}pin {E}cho, Phys. Rev. Lett. 55 (1985) 2370--2370.
\newblock \href {http://dx.doi.org/10.1103/PhysRevLett.55.2370.3}
  {\path{doi:10.1103/PhysRevLett.55.2370.3}}.

\bibitem{Pappas2017}
C.~Pappas, L.~J. Bannenberg, E.~Leli\`evre-Berna, F.~Qian, C.~D. Dewhurst,
  R.~M. Dalgliesh, D.~L. Schlagel, T.~A. Lograsso, P.~Falus, Magnetic
  {F}luctuations, {P}recursor {P}henomena, and {P}hase {T}ransition in {M}n{S}i
  under a {M}agnetic {F}ield, Phys. Rev. Lett. 119 (2017) 047203.
\newblock \href {http://dx.doi.org/10.1103/PhysRevLett.119.047203}
  {\path{doi:10.1103/PhysRevLett.119.047203}}.

\bibitem{Marry2013}
V.~Marry, E.~Dubois, N.~Malikova, J.~Breu, W.~Haussler, Anisotropy of {W}ater
  {D}ynamics in {C}lays: {I}nsights from {M}olecular {S}imulations for
  {E}xperimental {QENS} {A}nalysis, The Journal of Physical Chemistry C
  117~(29) (2013) 15106--15115.
\newblock \href {http://arxiv.org/abs/http://dx.doi.org/10.1021/jp403501h}
  {\path{arXiv:http://dx.doi.org/10.1021/jp403501h}}, \href
  {http://dx.doi.org/10.1021/jp403501h} {\path{doi:10.1021/jp403501h}}.

\end{thebibliography}

%%%%%%%%%%%%%%%%%%%%%%%%%%%%%%%%%%%%%%%%%%%%%%%%%%%%%%%%
%%%%%%%%%%%%%%%%%%%%%%%%%%%%%%%%%%%%%%%%%%%%%%%%%%%%%%%%
%%%%%%%%%%%%%%%%%%%%%%%%%%%%%%%%%%%%%%%%%%%%%%%%%%%%%%%%
%%%%%%%%%%%%%%%%%%%%%%%%%%%%%%%%%%%%%%%%%%%%%%%%%%%%%%%%
\end{document}